\documentclass[prl,preprint,floatfix]{revtex4}
\usepackage{epsfig}

\newcommand{\be}{\begin{equation}}
\newcommand{\ee}{\end{equation}}

\begin{document}

\title{The quark mass dependence of the pion mass at infinite $N$}
\author{R. Narayanan}
\affiliation{
Department of Physics, Florida International University, Miami,
FL 33199\\{\tt rajamani.narayanan@fiu.edu}}
\author{ H. Neuberger}
\affiliation{
Rutgers University, Department of Physics
and Astronomy,
Piscataway, NJ 08855\\{\tt neuberg@physics.rutgers.edu}
}

\begin{abstract}
In planar QCD, in two space time dimensions, the meson eigenvalue equation
has a nonlocal structure interpretable as resulting from hidden
degrees of freedom. The nonlocality can be reconstructed from 
the functional form of the pion mass dependence on quark mass within
an expansion starting from a special one dimensional Schr{\" o}dinger problem. 
The one dimensional problem makes the pion mass depend on the quark
mass through a simple quadratic relation which  is shown to be compatible 
also with numerical data obtained in four dimensions.
\end{abstract}

\maketitle

\section{Introduction}
Intuitively, if strings indeed describe QCD, the best 
place to make this precise 
is in the meson sector, at infinite number of colors, $N$. 
The mesons do not influence the gauge field vacuum and are open string probes
of the closed string background describing it. 
The mesons are free then, and, for
fixed quark mass (assume all quarks have equal mass for simplicity) their
masses squared should fall into some regular 
patterns, extending to infinite mass. 
An example is provided
by $SU(N)$ gauge theory at infinite $N$ in two space-time dimensions.
Although in two dimensions the gauge degrees of freedom are unlikely to provide a 
rich enough structure for a fully featured 
closed string background, a degenerate form, closer to
topological string theory, might exist~\cite{grosstaylor}.

A speculation~\cite{Polyakov} for four dimensions condenses
all the unknown structure into one real function of one real variable, 
playing the role of a warp factor in a five dimensional metric. 
This function determines the closed string background and the meson mass
dependence on quark mass. In simple examples  ~\cite{cvj} the warp factor directly
determines the masses of some particles by a local partial differential equation,
but in the case of QCD it is unlikely that the equation giving the meson masses
indeed is local in any set of variables. However, if one expands in energy,
relative to a scale set by the string tension, the mass of the lightest mesons
might be well described by a leading approximation, which does consist
of a simple second order partial differential equation. 
Higher orders in the two dimensional field theory describing the four dimensional QCD string
will induce further corrections in the meson-quark mass dependence. 
As we shall show below, in two dimensions, the function giving the
mass of the pion, $m_\pi^2$, as a function of the mass of the quark, $m_q$,
starts from a simple quadratic formula, structurally rooted in an ordinary second order eigenvalue
problem. The full, exact dependence then determines all the higher order corrections
in a low energy expansion which reproduces the full non-local meson equation. 
This leads us to a numerical test of a similar quadratic approximation in four dimensions.

The basic philosophy we adopt is to assume that mesonic spectral data could be used
to reconstruct the unknown free string theory purportedly describing planar QCD, somewhat
akin to an inverse scattering approach using spectral
and scattering data to get the potential in a Schr{\"o}dinger problem 
~\cite{invscat}. 

AdS/CFT motivated modeling seems to be relatively 
successful~\cite{models} phenomenologically, although
for four dimensional planar
QCD we only have ad-hoc motivated equations, and we even don't yet have 
the numerical values of 
the masses with their dependence on quark mass we should compare these to,
since the real world has only 3 colors.  As already mentioned, it is unlikely
that such equations, which work exactly in some AdS/CFT cases,  
at all exist as exact representations at infinite
$N$, and it is much less likely that any such equations are exact for $N=3$. 

In this letter we take some
steps to improve our understanding of mesonic planar QCD in two and four
dimensions by focusing on the dependence
of the lightest pseudo-Goldstone 
mesons (pions) on the quark mass at $N=\infty$.

Our work also provides a test of the numerical methodology
we have been developing for dealing with planar QCD in the meson sector.
Here, we shall skip most technical details, and concentrate on presenting
the results, speculating on their possible meaning. 

\section{Two dimensions}

We start in two dimensions, from 't Hooft's exact solution. The chiral
condensate $\langle \bar\psi\psi \rangle$ and the meson masses $m^2$ 
are exactly known as functions of $m_q$, the quark mass. The 't Hooft
coupling $\lambda=\frac{g^2 N}{\pi}$ is used to set the scale.
The physical dimensionless quantities are $\lambda^{-1/2} \langle \bar\psi\psi \rangle$, 
$\gamma=\lambda^{-1} m_q^2$ and $\mu^2 = \lambda^{-1} m^2$. 
Spontaneous symmetry breaking occurs
when the order of limits is: $\lim_{m_q\to 0} \lim_{N\to\infty} $, leaving a non-zero
condensate in the zero quark mass limit~\cite{zh}.

The meson spectrum and its dependence on quark mass (here
the quarks have been taken of equal mass for simplicity) is governed by 
't Hooft's equation~\cite{thooft}:
\begin{equation}
\label{thooft}
\gamma\left ( \frac{1}{x} +\frac{1}{1-x}\right ) \phi (x) - P\int_0^1 \frac{\phi(y)-\phi(x)}{(y-x)^2} dy =\mu^2\phi(x)
\end{equation}
$x$ is the fraction of meson light-cone momentum carried by one of the
quarks and varies between 0 and 1.

It is natural to ask whether 't Hooft's equation contains any structural
hints that it might admit a geometrical interpretation tied 
to self-consisted string propagation, which also could employ
extra dimensions for the string to propagate in. More precisely, as a first step, 
we wish to see whether the addition of some 
auxiliary continuous arguments can turn 't Hooft's equation into a local
differential equation of second order. The answer is positive.

While the first term in (\ref{thooft}) is local in $x$,
the second is highly non-local.  One can easily expand it in derivatives of $\phi$.
To deal with the combinatorics of the coefficients in a more efficient way, we seek a
set of functions that diagonalize this term. This set is easily found, using
the integral:
\begin{equation}
P\int_a^b dx \frac{(x-a)^{\nu-1}(b-x)^{-\nu}}{x-c} = -\frac{\pi(c-a)^{\nu-1}}{(b-c)^\nu}\cot (\nu\pi),
\end{equation}
where $a<c<b$ and $0<\Re\mu < 1$; we shall use the formula also at 
$\Re\nu =0,1$. 
It is now convenient to introduce $s=\log\frac{x}{1-x}$, measuring the
rapidity difference between the quark and anti-quark, and the equation
gets recast into the following form:
\begin{equation}
\left [ 1- \pi\frac{d}{ds} \cot \left ( \pi \frac{d}{ds} \right ) \right ] \Psi(s) +\frac{\mu^2}{4\cosh^2\frac{s}{2}}\Psi(s) =\gamma\Psi(s)
\end{equation}

In this form of the equation, 
$\mu^2$ is more naturally viewed as part of the potential 
term, while the eigenvalue role is more naturally taken up by $\gamma$, 
measuring the quark mass. Using
\begin{equation}
\label{n-exp}
\pi p \coth \pi p -1 = \sum_{n\ne 0} \frac{p^2}{p^2+n^2}
\end{equation}
it becomes obvious that we could localize the equation by 
adding a new dimension corresponding to the internal coordinate along
and open string, where the modes are labeled by nonzero integers $n$. 

Explicitly, we introduce a real field $\chi(s,\sigma)$, where
$\sigma\in [0,\pi]$, in addition to the field $\Psi(s)$, which
plays the role of the meson wave function, but is taken as real.
The field $\chi$ obeys Neumann boundary conditions in $\sigma$ and
is further constrained to contain no zero mode. 
\begin{equation}
\frac{\partial\chi(s,0)}{\partial\sigma} =\frac{\partial\chi(s,\pi)}{\partial\sigma}=0,~~~~
\int_0^\pi \chi(s,\sigma) =0
\end{equation}
We now introduce a local action functional $S[\Psi,\chi]$ whose
extremization subject to the above boundary condition produces the
meson wave equation.
\begin{eqnarray*}
&S[\phi,\chi]=\frac{1}{2}\int_{-\infty}^{\infty} ds \int_0^\pi
\frac{d\sigma}{\pi} \chi(s,\sigma) 
\left [  \frac{\partial^2}{\partial s^2} +\frac{\partial^2}{\partial\sigma^2}\right ]\chi(s,\sigma)-\\
&\int_{-\infty}^{\infty} ds\frac{d \Psi(s)}{ds} \chi(s,0)+\frac{1}{2}\int_{-\infty}^{\infty} ds \Psi(s) \left [
\gamma - \frac{\mu^2}{4\cosh^2(\frac{s}{2})} \right ] \Psi (s)
\end{eqnarray*}
\begin{equation}
\label{seq}
\chi(s,\sigma)=\sum_{n\ne 0} \cos(n\sigma)\chi_n (s)~~~\chi_n \equiv\chi_{-n}
\end{equation}
Eliminating the variables $\chi_n(s)$ reproduces the meson wave equation.
We have ended up with an open bosonic string that moves in one dimension
but whose center of mass is stuck. The nonlocality of (\ref{thooft}) 
reflects the integration over the stringy modes labelled by $n$ 
and the equations resemble those
of ~\cite{sjr}. There likely are other ways to localize the eigenvalue
equation, and it is unclear whether these various localizations
would be in some sense geometrically equivalent. 

The non-locality we have analyzed 
seems intrinsic, in the sense that no representation of the equation
is known which would make it local without the addition of 
extra independent variables.
There is one case when the non-locality can be neglected to a good
approximation:
When there is one very light particle, as a result of 
$\gamma << 1$, the nonlocality amounts to determining 
a parameter in a local description of the light particle, as one would expect.
We can approximate then (\ref{n-exp}) by assuming $p^2 << 1$ and 
then, in the light quark limit, one gets the usual form
\begin{equation}
\mu_\pi^2 (\gamma) = \frac{2\pi}{\sqrt{3}}\sqrt{\gamma} + {\cal O} (\gamma)
\end{equation}
The equation producing this result is a well known case
of an explicitly factorizable Schr{\" o}dinger equation, 
with a shape invariant potential~\cite{cooper}:
\begin{equation}
\label{cosheq}
\frac{\pi^2}{3}\frac{d^2}{ds^2}\Psi(s) +\frac{\mu^2}{4\cosh^2\frac{s}{2}}
\Psi(s)=\gamma\Psi(s) 
\end{equation}
This equation is an approximation, valid
for the lightest meson, in the limit of light quark mass. 
The roles of parameter and eigenvalue of meson mass and quark mass
got reversed.  The equation can be thought of as an analogue of
an AdS/CFT motivated partial differential equation determining the pion mass
for given quark mass, but only at leading order in an energy scale given by the
string tension. 
 
A systematic expansion in $(\frac{d}{ds})^2$ (this is not
a chiral Lagrangian expansion) allows the calculation of higher 
order terms in $\sqrt{\gamma}$. If we knew by some other means
the entire series in $\sqrt{\gamma}$ expressing $\mu_\pi^2 (\sqrt{\gamma}) $
we could iteratively 
invert this to determine the nonlocality of the differential
equation, once the lowest order~(\ref{cosheq}) is assumed as given. 
The exact function $\mu_\pi^2 (\sqrt{\gamma})$ reproduces all the sigma
model corrections to the leading form which here is given by a simple 
potential problem in one dimension. 

If we only know $\mu_\pi^2 (\sqrt{\gamma})$ numerically, we are limited
in the extend to which we can make inferences on the equation determining
the meson masses. This is the situation we find ourselves in four dimensions, 
where we don't have a general 
formula for the lowest order, starting point, equation either. 

All is not lost though, and one can try something much less ambitious than
determining the full equation for the mesons, something 
which nevertheless carries some  nontrivial information.  Let us see what
the approximate equations tells us about the function $\mu^2_\pi (\sqrt{\gamma})$.

That $\mu_\pi^2 (\sqrt{\gamma})$ vanishes linearly in $\sqrt{\gamma}$ as $\gamma\to 0$
follows from general field theoretical considerations. 
Because we are at infinite
$N$, all chiral logarithms are suppressed. By a simple calculation we obtain:
\begin{equation}
\label{twodchi}
\mu_\pi^2 (\gamma) = \frac{2\pi}{\sqrt{3}}\sqrt{\gamma} +4\left (1-\frac{\pi^2}{90}\right )\gamma +...
\end{equation}

Demanding the subleading correction to be smaller than $10\%$ of the leading
term gives
\begin{equation}
\frac{2\sqrt{\gamma}}{\mu_\pi} < \frac{1}{3}
\end{equation}

However, the more natural expansion of the full 't Hooft equation is in $(\frac{d}{ds})^2$.
The correction coming from the subleading term $\frac{d^4}{ds^4}$ is the $\frac{\pi^2}{90}$
term in (\ref{twodchi}), which is only ten percent of the order $\gamma$ correction. 
Had we used this as a correction when estimating the accuracy of (\ref{twodchi}) there
would have been no restriction, since the right hand side would have become unity, and this
is about as high as the ratio can ever get. Thus, one would say that (\ref{twodchi}) is uniformly valid to
ten percent accuracy. The factorization of~(\ref{cosheq}) into $-A^\dagger A$, with $A$ of first order in $\frac{d}{ds}$, induces one to replace $\mu^2$ by a variable $\Delta$:
\begin{equation}
\frac{1}{4} \mu^2 = \Delta (\Delta + \frac{\pi}{2\sqrt{3}})
\end{equation}
or,
\begin{equation}
\Delta=\frac{1}{2}\left  [ \sqrt{\mu^2 + \left ( \frac{\pi}{2\sqrt{3}}\right )^2} -  \frac{\pi}{2\sqrt{3}} \right ]
\end{equation}
Next, one expands $\Delta$ in $\sqrt{\gamma}$:
\begin{equation}
\Delta= \sqrt{\gamma} -\frac{\pi\sqrt{3}}{45}\gamma  ...
\end{equation}
Both expansions require the inclusion of the 
same amount of extra $(\frac{d}{ds})^{2n}$ terms at any fixed order, but the 
derivative expansion delivers reliable information about its accuracy while
the series expansion does not. Taking for example $\gamma=1$, and 
comparing (\ref{twodchi}) with the exact result, the approximation
is seen to be 
very good numerically, much better than the relative magnitude of the order
$\gamma$ term would have indicated, but in agreeemnt 
with the order $\frac{d^4}{ds^4}$ correction. 

The quadratic relation between the meson mass and the $\Delta$ variable
is reminiscent of calculations done in the AdS/CFT context, but the
meaning of $\Delta$ differs from the conformal case. In the
AdS/CFT case, the quadratic
nature of the relationship reflects its origin from an equation
that has no derivatives higher than second. This local structure
of the equation is a direct expression of the relevance of the extra (fifth for
four dimensional space-time) coordinate of AdS. 

The approximation in which we introduce $\Delta$ to replace $\mu^2$ and set 
$\Delta=\sqrt{\gamma}$ includes  
the correct leading asymptotic behavior of the quark mass dependence
of the pion both at very small and very large quark masses. However,
when we start expanding $\Delta$ in a power series in $\sqrt{\gamma}$, we are
only reproducing correctly subleading terms at the low quark mass end.
Using the exact equation in the nonlocal form given in (\ref{thooft}), 
it is possible
to set up an expansion of $\mu^2$ in $\gamma$ which is 
valid as both are very large.
To this end one needs to carry out a canonical change of variables, from
$(s,\frac{d}{ds})$ to $(\frac{d}{dq},-q)$. After that one can scale variables
so that the heavy quark limit is smooth and one obtains a leading form
of the wave equation that has a potential 
proportional to $|q|$ and a kinetic energy
of normal form.  The equation describes a point 
particle in a linear potential and
the fact that this holds in $q$, a variable 
conjugate to $s$, perhaps makes it easier to
accept the previous picture of an open string with Neumann boundary conditions
as describing a meson made out of massive quarks. 
From the new equation one extracts the following fact:
\begin{equation}
\lim_{\gamma\to\infty} \left [ \gamma-\frac{\mu^2}{4} -1 -
\left ( \frac{\mu\pi}{4}\right )^{2/3} \alpha \right ] =0
\end{equation}
$\alpha$ is the first zero of the derivative of the Airy function,
$\alpha\approx -1.0188$. As a result~\cite{negele}, 
\begin{equation}
\mu\approx 2\sqrt{\gamma} \left [1-\frac{\alpha}{2} 
\left ( \frac{\pi}{2} \right )^{2/3} \gamma^{-2/3} \right ], 
\end{equation}
a behavior that cannot be represented by a truncated power expansion of $\gamma$ in $\Delta$.
The expansion in the derivative terms $\left(  \frac{d^2}{ds^2}\right )^n$
is inappropriate at $m_q\to\infty$. At heavy quark masses, a 
different expansion, which produces the above result at 
leading order, can be used to generate a series of subleading terms.

From this discussion we extract the message that a parametrization of
the pion mass in terms of a quadratically related $\Delta$ 
produces an expansion of  the functional dependence of the pion mass 
on light quark mass which converges better. 
While the chiral expansion controls the structure at small quark masses,
what happens in the regime of intermediate quark masses already depends
on the specific dynamics of the model.  Because of he dynamics 
of this specific model, as opposed to other models with exactly the same
chiral Lagrangian to order $m_q^2$, a quadartic formula for $m_\pi^2$ in terms
of $m_q$ has a larger domain of applicability than can be argued on the basis
of the chiral Lagrangian alone.

Finally, this is something we can look for by numerical means.
It is not very restrictive, nor very solidly formulated, 
but testable in planar QCD in four space-time
dimensions.

\section{Numerical computation of the pion mass}

Recent work ~\cite{knn} has established that planar QCD on an Euclidean torus of size and shape
$l^4$ is $l$-independent so long as $l>l_c$. This is a string--like property. One refers to this
property as ``continuum reduction'', on account of the elimination of the infinite volume factor from 
the total number of degrees of freedom.
For a fixed bare gauge coupling, $g_0^2 N=\lambda_0$, this means that
computations in the large $N$ limit of four dimensional QCD can be
performed on an $L^4$ lattice of relatively small size, 
with $L > L_c(b)$, where $b=\frac{1}{\lambda_0}$.  
It is sufficient to pick $L$ just slightly above $L_c(b)$, 
since continuum reduction implies that there are no finite volume effects, once $N$ is large enough. 
One should think about $L$ as setting the minimal 
length scale in the problem, in this case 
given by $\frac{l_c}{L}$. 

The lattice gauge field configuration consists of a collection of $SU(N)$ matrices, associated
with a link in the direction $\hat\mu$ emanating from a site $x$ and denoted by $U_\mu(x)$.
These matrices are generated with a probability given by Wilson's plaquette action, with
coupling $b$, and the configuration is probed by a lattice fermion propagator that has
exact chiral symmetry at zero quark mass, known as the ``overlap''~\cite{overlap} propagator. 
Given the lattice gauge field  $U_\mu(x)$ on an $L^4$ lattice at some
coupling $b$, the lattice quark propagator using overlap fermions
is denoted by $G(U_\mu,m_o)$ where
\be
G(U_\mu,m_o) =  \frac{1}{1-m_o} \left [  
\frac{2}{1+m_o+(1-m_o)\gamma_5\epsilon[H_w(U_\mu)]}
-1 \right ]
\ee
and $m_o$ is the bare overlap quark mass parameter.  $H_w$ is the Wilson lattice Dirac
operator at mass $m_0$ with a so called $r$-parameter set to unity. 
The meson momentum is implemented by changing the gauge fields felt by the constituent
quarks by independent $U(1)$ phase factors; this is the so called ``quenched momentum prescription".
Meson propagators are computed using this quenched momentum prescription and are given by
\be
{\cal M}_\Gamma (p,m_o) = {\rm Tr} \left [
S \Gamma G(U_\mu e^{\frac{ip_\mu}{2}},m_o) 
S \Gamma^\dagger G(U_\mu e^{-\frac{ip_\mu}{2}},m_o)\label{meson}\right ]
\ee
The definitions of $S$ and $\Gamma$ follow below: 
We choose $\Gamma=\gamma_5$ for a pseudoscalar meson and
$\Gamma=1$ for a scalar meson.
The two quark propagators 
see gauge fields that differ by a $U(1)$ phase and this difference is the
momentum that is carried by the meson.
The meson momentum is taken along a lattice axis:
\be
p_\mu = \cases{ 0 & if $\mu=1,\cdots,3$ \cr \frac{2\pi n}{NL} & if $\mu=4$;  ~~~~~ $0 < n < N$}
\ee 
$S$ smears the operator in the remaining, perpendicular, 
$\mu=1,\cdots,d-1$ directions.
This ``smearing'' creates an extended object that more closely approximates the
true constituent quark structure of the meson. 
The smear operator is chosen to be the inverse of the gauged laplacian,
\be
S^{-1} = \frac{1}{2} \sum_{\mu=1}^{d-1} (2-T_\mu - T_\mu^\dagger) ,
\ee
where
$T_\mu$ is the gauge covariant translation operator defined by the
action $(T_\mu\psi)(x) = U_\mu(x)\psi(x+\hat\mu)$.  Because of fluctuations,
$S$ actually has a sizable gap and should be thought of as a lattice
version of the covariant Laplacian with a mass of the order of the inverse
lattice spacing.  
The lattice theory is in the confining, $l$-independent phase,  
when $L  > L_c(b)$; there the
$Z_N^4$ symmetries associated with the Polyakov loops in the four
directions are unbroken. Along with the gauge symmetry,  one can use this to show 
\be
{\cal M}_\Gamma (p,m_o) = {\rm Tr}  \left[ 
S \Gamma G(U_\mu e^{iq_\mu+\frac{ip_\mu}{2}},m_o) 
 S \Gamma^\dagger G(U_\mu e^{iq_\mu-\frac{ip_\mu}{2}},m_o)\right];
\ee
\be
q_\mu = \cases{ 0 & if $\mu=1,\cdots,3$ \cr \frac{2\pi n}{NL} & if $\mu=4$; ~~~
$0 < n < N$,}
\ee 
making it explicit that the meson propagator depends only on the difference between the
two phases seen by the two valence quarks.

Both the pseudoscalar meson and the scalar meson propagator computed above
are given by sums over an infinite number of poles. Smearing reduces the residues
associated with the higher poles. We can further reduce their contribution
by working with the sum, 
\be
F(p,m_o) = \frac{1}{2}\left [ {\cal M}_1(p,m_o)+{\cal M}_{\gamma_5}(p,m_o) \right ].
\ee
Excited pseudoscalar mesons and scalar mesons get closer in mass as
one looks at high excited levels in QCD~\cite{gloz} and one gets cancellations
between the higher states contributing to $F(p,m_0)$. In practice, a 
single pole fit to the gauge field average of $F(p,m_0)$ works well. 
\be
\langle F(p,m_o) \rangle = \frac {r_\pi^2(m_o)}{p^2 + m_\pi^2(m_o)}
\ee

We compute the meson propagator using a
stochastic estimate of the trace in (\ref{meson}).
We start by picking one Gaussian chiral source, $|q>$, such that
$\gamma_5 |q> = |q>$. Then we compute
$SG(e^{\frac{ip_\mu}{2}},m_o)|q>$ and 
$G(e^{-\frac{ip_\mu}{2}},m_o)S|q>$. 
The action of the overlap propagator
on a vector is computed using a standard multiple mass conjugate gradient algorithm.
The hermitian Wilson-Dirac operator, $H_w$, has a substantial spectral 
gap for large $N$ gauge fields due to the gap in the eigenvalue 
distribution of the single plaquette parallel transporter. This eliminates the need
to project out approximate zero modes from $H_w$, making the calculation straightforward. 
A satisfactorily accurate action of $\epsilon(H_w)$ is achieved
by just using the $21^{\rm st}$ order Zolotarev approximation.
The stochastic estimate of $F(p,m_0)$ is then given by
\be
\bar F(p,m_o) = <q| G^\dagger(e^{\frac{ip_\mu}{2}},m_o) S 
\frac{1+\gamma_5}{2} G(e^{-\frac{ip_\mu}{2}},m_o)S |q>
\ee
A multiple mass conjugate gradient algorithm for the calculation of $G$ easily traces
the dependence  on quark mass. Statistical errors are reduced by using the same
Gaussian chiral source all all momenta.
This numerical procedure results in an estimate for $\langle F(p,m_o)\rangle$
where the values at all $p$ and $m_o$ are correlated.

The stochastic estimate of $F(p,m_o)$ is then fitted as a function of $p$ at
a fixed quark mass to give $m_\pi^2(m_0)$. Minimization fits are performed
using the full correlation matrix and errors are estimated using the jackknife
method. We obtain in this way the pion mass as a function of 
the bare overlap quark
mass parameter at various gauge couplings. 
This result is free of finite volume effects since we
use large enough $N$, leaving only finite lattice spacing effects to worry about.

In order to take the continuum limit, we first need to convert the
bare overlap quark mass parameter to a physical quantity. As is well known,
the quark mass itself can be defined to renormalize multiplicatively, and with
the help of the overlap propagator this can be done also on the lattice. 
To eliminate this renormalization effect we consider instead 
the quantity $m_o\Sigma(b)$ where $\Sigma(b)$ is the
bare chiral condensate at a coupling of $b$ and zero quark mass. 
We can then convert both the pion
mass, $m_\pi$, and $m_o\Sigma(b)$ into dimensionless quantities 
by forming $m_\pi L_c(b)$ and $m_o \Sigma(b) L^4_c(b)$. Plots of
the functions relating these dimensionless quantities 
obtained at different lattice spacings
should approximately 
fall on a common curve that ceases to depend on the lattice spacing
as one approaches the continuum limit. 

\section{Numerical results}

\begin{table}
\begin{tabular}{|r|r|r|r|r|}
\hline
$b$ & $L$ & $N$ & $L_c(b)$ & $\Sigma^{1/3}(b)$ \\
\hline
0.345 & 8 & 19 & 4.77 & 0.1675\\
0.350 & 8 & 23 & 5.97 & 0.1420\\
0.355 & 10 & 19 & 6.96 & 0.1265\\
0.360 & 11 & 17 & 8.01 & 0.1130\\
\hline
\end{tabular}
\caption{ Simulation parameters, critical box size and bare chiral
condensate}
\label{tab1}
\end{table}
We work with four different couplings as shown in Table~\ref{tab1}.
The associated values for  $L$ and $N$ used in the numerical simulation 
are shown in the same table. The critical sizes, $L_c(b)$, are
known from our previous work on continuum reduction~\cite{knn} and
is given by
\begin{equation}
b_I = b e(b)~~~~~~~e(b)=\frac{1}{N} 
\langle Tr U_{\mu,\nu} (x) \rangle~~~~~
L_c(b)=0.26 \left ( \frac {11}{48\pi^2 b_I }\right )^{\frac{51}{121}} 
e^{\frac{24\pi^2 b_I}{11}}.
\end{equation}
We also have
a good estimate of the bare chiral condensate at $b=0.350$
from previous work~\cite{bpsipsi}, but the 
estimates at other couplings, like $b=0.355$, are not as accurate. 
Therefore, we use the known value
at $b=0.350$ and adjust the values at the other couplings such
that all plots of $m_\pi^2L_c^2(b)$ as a function of 
$m_o \Sigma(b) L_c^4(b)$, for different values of $b$, fall on one curve. This
did not have to happen, but it does, indicating 
that lattice spacing effects are below our statistical errors.
The values of $\Sigma(b)$ so obtained are shown in Table~\ref{tab1}.
The value at $b=0.355$ in Table~\ref{tab1}
is higher by less than $5\%$ compared
to the numbers in~\cite{bpsipsi} but this is well within the error.
The value at $b=0.345$ 
used here is compatible with the value found at $b=0.346$ in \cite{bpsipsi}
The value at $b=0.360$ used here indicates
that the value at $b=0.3585$ found in \cite{bpsipsi} probably could
go up a little bit if one does a careful analysis.
There are potential order $a^2$ differences between the condensate
values listed in Table~\ref{tab1} and those determined 
using chRMT in \cite{bpsipsi}.
The running of $\Sigma(b)$ is given up to one-loop by~\cite{peskin}
\begin{equation}
\Sigma(b) L_c^3(b) 
= C \left[ \ln (L_c(b)\Lambda) \right]^{\frac{9}{22}}
\end{equation}
We find that $C=0.828$ and $\Lambda=0.268$ fits the numbers in Table~\ref{tab1}
quite well for $b\ne 0.345$ and we find a $8\%$ deviation at $b=0.345$.

\begin{figure}
\epsfxsize = 0.95\textwidth
\centerline{\epsfbox{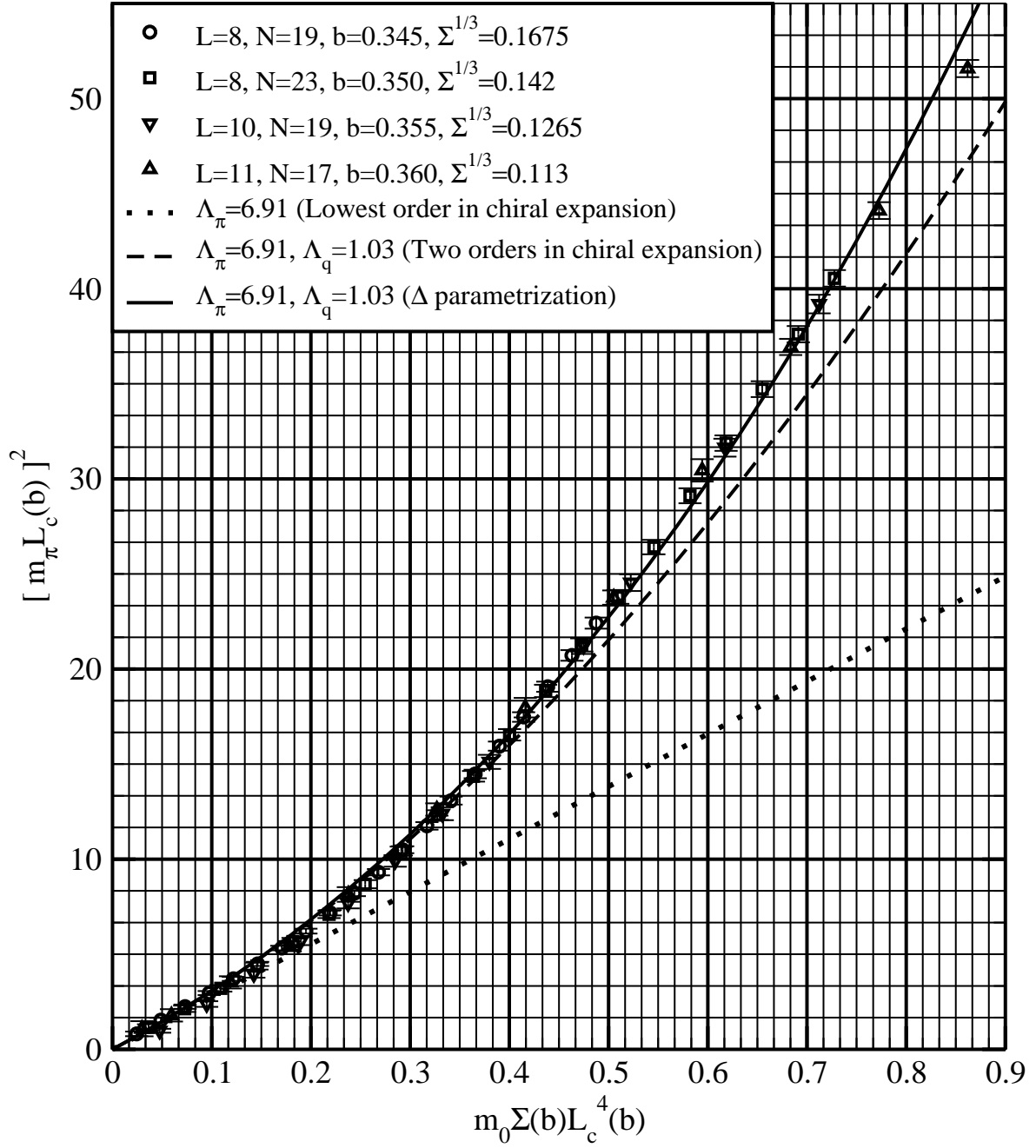}}
\caption{A plot of pion mass as a function of quark mass in
dimensionless units. 
}
\label{pion}
\end{figure}

The plot of $m_\pi^2L_c^2(b)$ as a function of 
$m_o \Sigma(b) L_c^4(b)$ is shown in Fig.~\ref{pion}.
The data is then fitted to
\begin{equation}
\Delta=\frac{1}{2} [\sqrt{m_\pi^2 L_c^2(b) + \Lambda_\pi^2}-\Lambda_\pi];~~
\frac{1}{4} m_\pi^2L_c^2(b)=\Delta(\Delta+\Lambda_\pi )
\end{equation}
with
\begin{equation}
\Delta= m_o\Sigma(b)L_c^4(b) +\frac{1}{\Lambda_q} m_o^2\Sigma^2(b)L_c^8(b) +....
\end{equation}
$\Lambda_\pi=6.91$ and $\Lambda_q=1.03$
fit the data over the whole range as shown by the solid
black curve in Fig.~\ref{pion}. A plot of a truncated chiral expansion,
\be
m_\pi^2 L_c^2(b) = 4\Lambda_\pi m_o\Sigma(b)L_c^4(b) +
4(1+\frac{\Lambda_\pi}{\Lambda_q}) m_o^2\Sigma^2(b)L_c^8(b),
\ee
shows that the second term in the above equation makes a large
contribution for $m_o\Sigma(b)L_c^4(b) > 0.2$. Keeping two orders
in the chiral expansion also does not agree with the data as well
as the $\Delta$ parametrization. One should note that 
$\Delta >> \Lambda_\pi$ 
in the large quark mass limit and this is not the case for the
range plotted in Fig.~\ref{pion}.

The parameter $\Lambda_\pi$ governs the low quark mass regime through
the relation
$f_\pi = \frac{1}{\sqrt{2\Lambda_\pi}l_c}$.
Using, $1/l_c = T_c = 264$~MeV, we get $f_\pi=71$~MeV.
This translates to $f_\pi=123$~MeV for SU(3) at zero quark mass, significantly 
 higher than the conventional value of 86~MeV.  Apparently, $\frac{1}{N}$ effects
 on the pion decay constant are larger than on the glueball mass or the finite
 temperature phase transition, all measured in units of string tension.

\section{Summary}

For simple quantum mechanical problems the spectrum and scattering data
can be used to determine the potential by the inverse scattering method.
Could we use spectral meson data at infinite $N$ to determine the string
descriptions of mesons ?  

In this paper we took a very primitive first step with this philosophy in
mind. We were led to 
the speculation that the structure of QCD in the
planar limit, where chiral logarithms are suppressed, explains why
mass formulae of the type $m_\pi^2 = 2B m_q + c m_q^2$ seem to work even
when $\frac{c m_q}{2B}$ is not small.  Of course, this may
be just saying  that the heavy quark regime, where $m_\pi\sim 2m_q$ is smoothly
connected to the light quark regime. We observed that this could also
reflect the approximate validity of an equation similar to~(\ref{cosheq}).

\section{Acknowledgements}

R. N. acknowledges partial support by the NSF under
grant number PHY-0300065 and also partial support from Jefferson 
Lab. The Thomas Jefferson National Accelerator Facility
(Jefferson Lab) is operated by the Southeastern Universities Research
Association (SURA) under DOE contract DE-AC05-84ER40150.
H. N. acknowledges partial support 
by the DOE under grant number 
DE-FG02-01ER41165 at Rutgers University, discussions with R. Brower, 
M. Einhorn, D. Gross, A. Jevicki, J. Polchinski, S-J Rey and A. Zhitnitsky,
and thanks the KITP for hospitality.

\end{document}